\documentclass[twoside,english,american,prb,floatfix,showpacs,preprintnumbers,superscriptaddress,amsmath,amssymb]{revtex4}
\usepackage{amsmath}
\usepackage{graphicx}
\usepackage{amssymb}

\makeatletter
\usepackage{bbm}

\usepackage{babel}
\makeatother
\begin{document}
\selectlanguage{english}

\title{Cluster-based density-functional approach to quantum transport \\
through molecular and atomic contacts}

\selectlanguage{american}

\author{F.~Pauly}

\email{fabian.pauly@kit.edu}

\affiliation{Institut f\"ur Theoretische Festk\"orperphysik and DFG-Center for
Functional Nanostructures, Universit\"at Karlsruhe, 76128 Karlsruhe,
Germany}

\affiliation{Institut f\"ur Nanotechnologie, Forschungszentrum Karlsruhe, 76021
Karlsruhe, Germany}

\author{J.~K.~Viljas}

\affiliation{Institut f\"ur Theoretische Festk\"orperphysik and DFG-Center for
Functional Nanostructures, Universit\"at Karlsruhe, 76128 Karlsruhe,
Germany}

\affiliation{Institut f\"ur Nanotechnologie, Forschungszentrum Karlsruhe, 76021
Karlsruhe, Germany}

\author{U.~Huniar}

\selectlanguage{english}

\affiliation{COSMOlogic GmbH \& Co.~KG, Burscheider Str.~515, 51381 Leverkusen,
Germany}

\selectlanguage{american}

\affiliation{Institut f\"ur Physikalische Chemie, Universit\"at Karlsruhe, 76128
Karlsruhe, Germany}

\author{M.~H\"afner}

\affiliation{Institut f\"ur Theoretische Festk\"orperphysik and DFG-Center for
Functional Nanostructures, Universit\"at Karlsruhe, 76128 Karlsruhe,
Germany}

\author{S.~Wohlthat}

\affiliation{Institut f\"ur Theoretische Festk\"orperphysik and DFG-Center for
Functional Nanostructures, Universit\"at Karlsruhe, 76128 Karlsruhe,
Germany}

\affiliation{School of Chemistry, The University of Sydney, Sydney, NSW 2006,
Australia}

\author{M.~B\"urkle}

\affiliation{Institut f\"ur Theoretische Festk\"orperphysik and DFG-Center for
Functional Nanostructures, Universit\"at Karlsruhe, 76128 Karlsruhe,
Germany}

\author{J.~C.~Cuevas}

\affiliation{Departamento de F\'isica Te\'orica de la Materia Condensada, Universidad
Aut\'onoma de Madrid, 28049 Madrid, Spain}

\affiliation{Institut f\"ur Theoretische Festk\"orperphysik and DFG-Center for
Functional Nanostructures, Universit\"at Karlsruhe, 76128 Karlsruhe,
Germany}

\affiliation{Institut f\"ur Nanotechnologie, Forschungszentrum Karlsruhe, 76021
Karlsruhe, Germany}

\author{Gerd Sch\"on}

\affiliation{Institut f\"ur Theoretische Festk\"orperphysik and DFG-Center for
Functional Nanostructures, Universit\"at Karlsruhe, 76128 Karlsruhe,
Germany}

\affiliation{Institut f\"ur Nanotechnologie, Forschungszentrum Karlsruhe, 76021
Karlsruhe, Germany}

\date{\today}

\begin{abstract}
We present a cluster-based density-functional approach to model charge
transport through molecular and atomic contacts. The electronic structure
of the contacts is determined in the framework of density functional
theory, and the parameters needed to describe transport are extracted
from finite clusters. A similar procedure, restricted to nearest-neighbor
interactions in the electrodes, has been presented by \foreignlanguage{english}{Damle
\emph{et al.}~{[}Chem.~Phys.~\textbf{281}, 171 (2002)]}. Here,
we show how to systematically improve the description of the electrodes
by extracting bulk parameters from sufficiently large metal clusters.
In this way we avoid problems arising from the use of nonorthogonal
basis functions. For demonstration we apply our method to electron
transport through Au contacts with various atomic-chain configurations
and to a single-atom contact of Al.
\end{abstract}


\maketitle
\selectlanguage{english}

\section{Introduction}

\selectlanguage{american}
Advances in the experimental techniques for manipulating and contacting
atomic-sized objects have turned the vision of molecular-scale electronic
circuits into a realistic goal.\foreignlanguage{english}{\cite{Nazin:Science2003,Nitzan:Science2003,Tao:NatureNano2006,Lindsay:AdvMater2007,Akkerman:JPhysCondMat2008}}
This has intensified the interdisciplinary efforts to study charge
transport in nanostructures. Ideally, the circuits would be constructed
in a bottom-up approach with functional units and all the wiring on
the molecular scale. To approach the goal, present-day experiments
in the area of molecular electronics concentrate on measuring the
current-voltage response of single molecules in contact to metallic
electrodes. In these studies, also purely metallic atomic contacts
and wires serve as important reference systems.\foreignlanguage{english}{\cite{Agrait:PhysRep2003}}

\selectlanguage{english}
In order to support the experiments and to stimulate further technological
advance, theoretical modeling of atomic-scale charge transport is
needed. Here one faces the challenge to describe infinitely extended,
low-symmetry quantum systems that may, in addition, be far from equilibrium
and involve strong electronic correlations. While a complete theoretical
understanding is still lacking, sophisticated \emph{ab-initio} methods
have been developed for approximate but parameter-free numerical simulations.
In order to study the prototypical metal-molecule-metal systems or
metallic atomic contacts, many groups use density functional theory
(DFT) combined with nonequilibrium Green's function (NEGF) techniques.\cite{Taylor:PRB2001,Xue:JCP2001,Damle:PRB2001,Brandbyge:PRB2002,Palacios:PRB2002,Heurich:PRL2002,Fujimoto:PRB2003,Wortmann:PRB2002b,Jelinek:PRB2003,Havu:PRB2004,Tada:JCP2004,Evers:PRB2004,Khomyakov:PRB2004,Calzolari:PRB2004,Thygesen:ChemPhys2005,Rocha:PRB2006}
Some shortcomings related to the use of DFT in this context have been
pointed out, and solutions are being sought.\cite{Feretti:PRL2005,Toher:PRL2005,Darancet:PRB2007,Thygesen:PRB2008}
On the other hand, from a practical point of view DFT presently appears
to be one of the most useful \emph{ab-initio} electronic structure
methods, since studies of quantum transport require dealing with a
large number of atoms. Furthermore the metal-molecule-metal contacts
are hybrid systems, where the central regions frequently behave rather
insulator-like, while the electrodes are metallic. For more complete
discussions we refer to Refs.~\onlinecite{Cuevas:Nanotec2003,Pecchia:RepProgPhys2004,Koentopp:JPCM2008}.

\selectlanguage{american}
The DFT approaches can mainly be divided into two types. In the first
one, atomic-sized contacts are modelled by periodically repeated supercells,
and computer codes developed for solid-state calculations are employed.\foreignlanguage{english}{\cite{Brandbyge:PRB2002,Thygesen:ChemPhys2005,Rocha:PRB2006}}
The use of periodic boundary conditions facilitates the electrode
description. However, the conductance is determined for an array of
parallel junctions and may be affected by artifical interactions between
them. The second type is based on finite clusters and originates more
from the chemistry community.\foreignlanguage{english}{\cite{Damle:PRB2001,Xue:JCP2001,Palacios:PRB2002,Heurich:PRL2002,Tada:JCP2004}}
It has the advantage that genuinely single-atom or single-molecule
contacts are described, and it makes possible investigations of molecules
of large transverse extent. The drawback is typically the description
of the electrode, since it is difficult to treat bulk properties based
on finite clusters. Furthermore the coupling between the device region
and the electrode can be complicated by finite-size and surface mismatch
effects.

To arrive at an \foreignlanguage{english}{\emph{ab-initio}} DFT description
it is necessary to treat the whole system consistently by using the
same basis set and exchange-correlation functional everywhere. The
problem of the cluster-based approaches regarding the electrodes is
apparent, for example, from the work of Refs.\foreignlanguage{english}{~\onlinecite{Xue:JCP2001,Palacios:PRB2002,Heurich:PRL2002}},
where the authors resort to a separate tight-binding parameterization
obtained from the literature.\foreignlanguage{english}{\cite{Papa:Book1986}}
Damle \foreignlanguage{english}{\emph{et al.}~}proposed to resolve
this issue by extracting electrode parameters from finite clusters
computed within DFT.\foreignlanguage{english}{\cite{Damle:PRB2001,Damle:ChemPhys2002}}
However, their treatment of the electrodes should be seen as a first
approximation, since only couplings between nearest neighbor atoms
were considered. Furthermore, they finally use energy-independent
self-energies, which is well-justified only for electrode materials
with a constant density of states (DOS) near the Fermi energy.

In this work we present a cluster-based DFT approach for the atomistic
description of quantum transport. We follow the ideas of Ref.~\foreignlanguage{english}{\onlinecite{Damle:ChemPhys2002}},
but place special emphasis on the treatment of the electrodes. In
particular, we show that extracting electrode parameters from small
metal clusters can lead to an unphysical behavior of the overlap of
the nonorthogonal basis functions in $k$-space. The description of
the electrodes can be improved systematically by employing metal clusters
of increasing size. Our implemenation is based on the quantum-chemistry
package TURBOMOLE, which allows us to treat clusters of several hundred
atoms. In this way we obtain an \emph{ab-initio} formulation of quantum
transport in atomic-sized contacts, where the whole system is treated
on an equal footing. It has the advantage that we can employ high-quality
quantum-chemical Gaussian basis sets, which are well-tested for isolated
systems.

\selectlanguage{english}
The theoretical framework of our approach is presented in Sec.~\ref{sec:Theoretical-approach}.
Several technical details, related to the use of nonorthogonal basis
functions and the electrode treatment can be found in Apps.~\ref{sec:NO-basis-sets}
and \ref{sec:electrodeDFT-electrode-description}. To demonstrate
the power of our methods we study in Sec.~\ref{sec:AuAlcontacts}
the transport properties of atomic contacts of Au and Al. The choice
of these materials is motivated by the fact that Au exhibits a rather
energy-independent DOS near the Fermi energy, while Al does not. Furthermore,
for these systems we can compare our results to the literature. We
find good agreement, and demonstrate the robustness of our results.
Further applications have been presented in Refs.~\onlinecite{Wohlthat:PRB2007,Viljas:PRB2007b,Wohlthat:ChemPhysLett2008,Pauly:PRB2008,Pauly:arXiv:0709}.
We summarize our results in Sec.~\ref{sec:Conclusions}.

\section{Theoretical approach\label{sec:Theoretical-approach}}

\subsection{Electronic structure\label{sub:Electronic-structure}}

\selectlanguage{american}
Our \emph{ab-initio} calculations are based on the implementation
of DFT in TURBOMOLE 5.9.\cite{Ahlrichs:ChemPhysLett1989} By \emph{ab-initio}
we mean that the simulations require no system-specific parameters.
Self-consistent DFT calculations of large systems are generally very
time-consuming. TURBOMOLE, however, is specialized in handling such
systems, and offers several possibilities to reduce the computational
effort. Thus, one can exploit point group symmetries, including non-Abelian
ones. In this way, the calculations speed up by a factor given by
the order of the point group. Further options are the {}``resolution
of the identity in $J$'' (RI-$J$)\cite{Eichkorn:ChemPhysLett1995,Eichkorn:TheorChimAcc1997}
and the {}``multipole-accelerated resolution of the identity $J$''
(MARI-$J$),\cite{Sierka:JCP2003} which are both implemented in the
ridft module of TURBOMOLE. The approximations help to reduce the effort
to compute the Coulomb integrals $J$, which are particularly expensive
to evaluate. With the help of the RI-$J$ approximation, known also
under the name {}``density fitting'', the four-center-two-electron
integrals can be expressed as three-center-two-electron ones.\cite{Koch:Wiley2001}
Calculations are faster by a factor of 10-100 as compared to standard
DFT, but equally accurate. The MARI-$J$ technique concerns the Coulomb
interactions between distant atoms. They are divided into a near-field
and a far-field part, where the near field is treated with RI-$J$
and the far field by a multipole approximation. Compared to RI-$J$,
it can accelerate the calculations by another factor of 2-7.\cite{Sierka:JCP2003}
DFT requires the choice of an exchange-correlation functional.\cite{Fiolhais:Springer2003}
We select the generalized-gradient functional BP86,\cite{Becke:PRA1988,Perdew:PRB1986}
which is known to yield good results for large metal clusters.\cite{Ahlrichs:PCCP1999,Furche:JCP2002,Koehn:PCCP2001,Nava:PCCP2003}
To express the orbital wave functions, Gaussian basis sets of split-valence-plus-polarization
(SVP) quality are used,\cite{Schaefer:JCP1992,Eichkorn:ChemPhysLett1995,Eichkorn:TheorChimAcc1997}
which are the TURBOMOLE standard. Within the closed-shell formalism
of DFT, total energies of all our clusters are converged to a precision
better than $10^{-6}$ a.u. In order to obtain ground-state structures,
the total energy needs to be minimized with respect to the nuclear
coordinates. We perform such geometry optimizations or {}``relaxations''
until the maximum norm of the Cartesian gradient has fallen below
$10^{-4}$ a.u.

\selectlanguage{english}

\subsection{Transport formalism\label{sub:Transport-formalism}}

\selectlanguage{american}
We compute transport properties of atomic-sized contacts using the
Landauer-B\"uttiker theory and Green's functions expressed in a local,
nonorthogonal basis.\cite{Xue:ChemPhys2002,Viljas:PRB2005} The local
atomic basis allows us to partition the basis states $\left|i,\alpha\right\rangle $
into left ($L$), central ($C$), and right ($R$) ones, according
to a division of the contact geometry. In the basis states, $\alpha$
refers to the type of orbital at the position of atom $i$. For reasons
of brevity we will frequently suppress the orbital index. The Hamiltonian
(or Kohn-Sham) matrix $H_{i\alpha,j\beta}=\left\langle i,\alpha\right|H\left|j,\beta\right\rangle $,
and analogously the overlap matrix $S_{i\alpha,j\beta}=\left\langle i,\alpha\right.\left|j,\beta\right\rangle $,
can thus be written in block form,\begin{equation}
H=\left(\begin{array}{ccc}
H_{LL} & H_{LC} & 0\\
H_{CL} & H_{CC} & H_{CR}\\
0 & H_{RC} & H_{RR}\end{array}\right).\label{eq:Hamiltonian}\end{equation}
Both $S$ and $H$ are real-valued and are hence symmetric. In addition,
we assume the $C$ region to be large enough to have $S_{LR}=H_{LR}=0$.
Within the Landauer-B\"uttiker theory,\cite{Datta:Cambridge2005}
the linear conductance can be expressed as\begin{equation}
G=\frac{2e^{2}}{h}\int dE\left[-\frac{\partial}{\partial E}f(E,T)\right]\tau(E),\label{eq:conductance_Tfinite}\end{equation}
where $f(E,T)=\left\{ \exp\left[\left(E-\mu\right)/\left(k_{B}T\right)\right]+1\right\} ^{-1}$
is the Fermi function, and the chemical potential $\mu$ is approximately
equal to the Fermi energy $E_{F}$, $\mu\approx E_{F}$. Using the
standard NEGF technique, the transmission function is given by \begin{equation}
\tau(E)=\mathrm{Tr}\left[\Gamma_{L}(E)G_{CC}^{r}(E)\Gamma_{R}(E)G_{CC}^{a}(E)\right]=\mathrm{Tr}\left[t^{\dagger}(E)t(E)\right]\label{eq:T_E_GammaGGammaG_ttd}\end{equation}
with the transmission matrix \begin{equation}
t(E)=\sqrt{\Gamma_{R}(E)}G_{CC}^{a}(E)\sqrt{\Gamma_{L}(E)}.\label{eq:transmission_matrix}\end{equation}
Here we define the Green's functions\begin{equation}
G_{CC}^{r}(E)=\left[ES_{CC}-H_{CC}-\Sigma_{L}^{r}(E)-\Sigma_{R}^{r}(E)\right]^{-1}\label{eq:GrCC}\end{equation}
and $G_{CC}^{a}=\left[G_{CC}^{r}\right]^{\dagger}$, the self-energies\begin{equation}
\Sigma_{X}^{r}(E)=\left(H_{CX}-ES_{CX}\right)g_{XX}^{r}(E)\left(H_{XC}-ES_{XC}\right),\label{eq:Simga_X}\end{equation}
and the scattering-rate matrices \begin{equation}
\Gamma_{X}(E)=-2\mathrm{Im}\left[\Sigma_{X}^{r}(E)\right],\label{eq:Gamma_X}\end{equation}
where $g_{XX}^{r}$ is the electrode Green's function for lead $X=L$,$R$.
At low temperatures, the expression for the conductance simplifies
to\begin{equation}
G=\frac{2e^{2}}{h}\tau(E_{F})=G_{0}\sum_{n}\tau_{n}(E_{F}),\label{eq:conductance_T0}\end{equation}
with $G_{0}=2e^{2}/h$ the quantum of conductance and $\tau_{n}$
the eigenvalues of $t^{\dagger}t$. The latter are the transmission
probabilities of the transmission channels $n$.%
\footnote{\selectlanguage{english}
The fact that the $\tau_{n}$ are probabilities, i.e.~that $0\leq\tau_{n}\leq1$,
can be proven by defining $r=\mathbbm{1}+i\sqrt{\Gamma_{L}}G_{CC}^{a}\sqrt{\Gamma_{L}}$
such that $r^{\dagger}r+t^{\dagger}t=\mathbbm{1}$. \foreignlanguage{american}{It
is easy to show the positive-semidefiniteness of $\Gamma_{L}$, necessary
for the determination of both $r$ and $t$ {[}Eq.~(\ref{eq:transmission_matrix})].}\selectlanguage{american}
} Also other observables, such as the thermopower or the photoconductance,
can be studied based on the knowledge of $\tau(E)$.\cite{Viljas:2007a,Viljas:PRB2007b,Pauly:arXiv:0709,Viljas:PRB2008}

\selectlanguage{english}
Information on the energetics of a system may help to identify conduction
mechanisms. Such information can be extracted from the spectral density\cite{Economou:Springer2006}
\begin{equation}
\rho(E)=\frac{i}{2\pi}\left[G^{r}(E)-G^{a}(E)\right]=-\frac{1}{\pi}\mathrm{Im}\left[G^{r}(E)\right].\label{eq:rhoEdef}\end{equation}
Using this, we define the local density of states (LDOS) at atom $i$
and its decomposition into orbitals $\alpha$ via\begin{eqnarray}
\mathrm{LDOS}_{i}(E) & = & \sum_{\alpha}\mathrm{LDOS}_{i\alpha}(E)\label{eq:LDOS_i}\\
\mathrm{LDOS}_{i\alpha}(E) & = & \left(S_{CC}^{1/2}\varrho_{CC}(E)S_{CC}^{1/2}\right)_{i\alpha,i\alpha}\label{eq:LDOS_ia}\end{eqnarray}
\foreignlanguage{american}{In App.~\ref{sec:NO-basis-sets} we discuss
the approximations involved in this definition. There we also consider
further the issues related to the use of nonorthogonal basis sets
to evaluate the single-particle Green's functions and the electric
current.}

\selectlanguage{american}

\subsection{Implementation of the transport method}

\subsubsection{Central system}

\begin{figure}
\selectlanguage{english}
\begin{centering}\includegraphics[width=0.5\columnwidth,clip=]{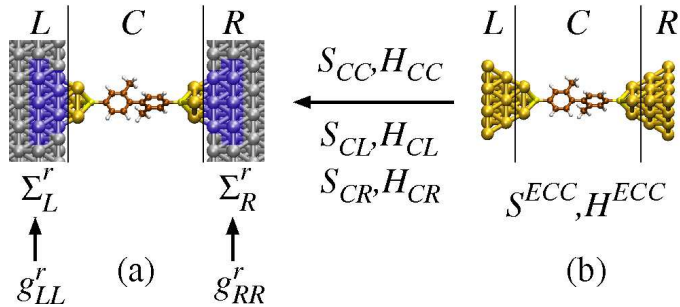}\par\end{centering}

\caption{\label{cap:QTscheme}Quantum transport scheme. The conduction properties
of an atomic-sized contact (a) shall be studied. For this purpose
it is divided into a $C$ region and two semi-infinite $L$ and $R$
electrodes. Using a similar division as for the contact, information
on the electronic structure of the $C$ region ($S_{CC}$, $H_{CC}$)
as well as the $CL$ and $CR$ couplings ($S_{CL}$, $H_{CL}$ and
$S_{CR}$, $H_{CR}$) is extracted from the ECC (b). In order to obtain
the self-energies $\Sigma_{L}^{r}$ \foreignlanguage{american}{and}
$\Sigma_{R}^{r}$ (a), the remaining task is to determine the electrode
surface Green's functions $g_{LL}^{r}$ and $g_{RR}^{r}$. This procedure
is described further below in the text.}\selectlanguage{american}

\end{figure}

In order to determine the transmission function $\tau(E)$ we need
a practical scheme to obtain the necessary information on the electronic
structure. In Fig.~\ref{cap:QTscheme} we present our procedure.%
 The goal is to describe the whole atomic-sized contact {[}Fig.~\ref{cap:QTscheme}(a)]
consistently, by treating the $L$, $C$, and $R$ regions with the
same basis set and exchange-correlation functional. We obtain the
parameters $S_{CC}$ and $H_{CC}$ as well as the couplings to the
electrodes $S_{CX}$ and $H_{CX}$ with $X=L,R$ from the extended
central system (ECC) {[}Fig.~\ref{cap:QTscheme}(b)], in which we
include large parts of the tips of the metallic electrodes. The division
of the ECC into the $L$, $C$, and $R$ regions is performed so that
the $C$ region is identical to that in Fig.~\ref{cap:QTscheme}(a).
The atoms in the $L$ and $R$ parts of the ECC {[}blue-shaded regions
in Fig.~\ref{cap:QTscheme}(a)] correspond to that part of the electrodes
that is assumed to couple to $C$. The partitioning or division of
the ECC is commonly made somewhere in the middle of the metal tips,
and we will also refer to it as \char`\"{}cut\char`\"{}. The electrodes
{[}$L$ and $R$ regions in Fig.~\ref{cap:QTscheme}(a)] are modeled
as surfaces of semi-infinite crystals, described by the surface Green's
functions $g_{XX}^{r}$. They are constructed from bulk parameters,
extracted from large metal clusters. Further below we discuss in detail,
how this is accomplished.

In our approach, we assume the metal tips included in the ECC to be
large enough to satisfy basically two criteria. First, all the charge
transfer between the $L$ and $R$ electrodes and the $C$ part of
the contact should be accounted for. This ensures the proper alignment
of the electronic levels in $C$ with $E_{F}$. Second, most of the
metal tips, especially the $L$ and $R$ parts of the ECC, should
resemble bulk as closely as possible. In this way, we can evaluate
the surface Green's functions by using bulk parameters of an infinitely
extended crystal. Owing to surface effects caused by the finite size
of the ECC, this can be satisfied only approximately. The mismatch
between the parameters in the $L$ and $R$ regions of the contact
and the ECC will thus lead to spurious scattering at the $LC$ and
$CR$ interfaces. In principle this resistance can be eliminated systematically
by including more atoms in the metal tips of the ECC. On the other
hand, if the resistance in the $C$ region is much larger than the
spurious $LC$ and $CR$ interface resistances, they will have little
influence on the results.

\selectlanguage{english}

\subsubsection{Electrodes}

\begin{figure}
\selectlanguage{english}
\begin{centering}\includegraphics[width=0.6\columnwidth,clip=]{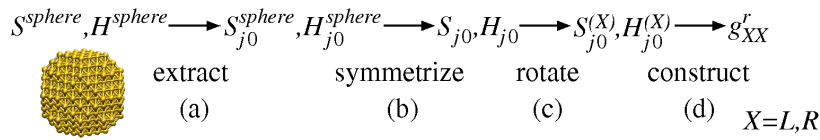}\par\end{centering}

\selectlanguage{american}

\caption{\label{cap:DFT-electrode-construction}\foreignlanguage{english}{In
order to obtain the electrode Green's functions $g_{XX}^{r}$ for
lead $X=L,R$, we determine bulk parameters from large metal clusters.
In a first step (a) we extract overlap and hopping elements, $S_{j0}^{sphere}$,
$H_{j0}^{sphere}$, from the cluster's central atom to all its neighbors.
They are (b) symmetrized by imposing the space-group of the electrode
lattice. After (c) a rotation to adapt them to the orientation of
the respective electrode, (d) $g_{XX}^{r}$ is constructed with the
help of a decimation procedure.}}
\end{figure}

\selectlanguage{american}
We extract bulk parameters describing perfect crystals from large
metal clusters. The complete procedure, which aims at determining
the surface Green's functions $g_{XX}^{r}$ with $X=L,R$ is summarized
in Fig.~\foreignlanguage{english}{\ref{cap:DFT-electrode-construction}}.
In this work we study exclusively electrode materials with an fcc
structure, of which Au and Al are examples.

In a first step {[}Fig.~\foreignlanguage{english}{\ref{cap:DFT-electrode-construction}}(a)]
we construct spherical metal clusters, henceforth called {}``spheres''.
They are made up of atoms at positions $\{\vec{R}_{j}|\vec{R}_{j}=\sum_{n=1}^{3}j_{n}\vec{a}_{n}\wedge|\vec{R}_{j}|\leq R^{sphere}\}$
with the standard fcc primitive lattice vectors $\vec{a}_{n}$ and
the sphere radius $R^{sphere}$. Here, we will use the vector of integer
indices $j=\left(j_{1},j_{2},j_{3}\right)$ to characterize the atomic
position $\vec{R}_{j}$. We do not relax the spheres, but set the
lattice constant $a_{0}$ to its experimental literature value.\cite{Ashcroft:Harcourt1976}
Increasing the radius $R^{sphere}$ should make the electronic structure
in the center resemble that of a crystal. From the clusters we extract
the overlap and Hamiltonian between the central atom at position $0$
and the neighboring ones at position $j$ (including $j=0$). This
yields the matrix elements $S_{j\alpha,0\beta}^{sphere}$ and $H_{j\alpha,0\beta}^{sphere}$,
where $\alpha$ and $\beta$ stand for the basis functions of the
atoms at $j$ and $0$. For reasons of brevity, we will often suppress
the orbital indices. $S_{j0}^{sphere}$ and $H_{j0}^{sphere}$ are
then matrices with appropriate dimensions.

The bulk parameters $S_{j0}$, $H_{j0}$ need to satisfy the symmetries
of the fcc space-group {[}Fig.~\foreignlanguage{english}{\ref{cap:DFT-electrode-construction}}(b)].
While $S_{j0}^{sphere}$ depends only on the relative position of
atoms, surface effects due to the finite size of the fcc clusters
lead to deviations from the translational symmetry for $H_{j0}^{sphere}$.
A rotation may still be necessary to arrive at parameters $S_{j0}^{\left(X\right)}$,
$H_{j0}^{\left(X\right)}$, which are adapted to the orientation of
the electrode $X=L,R$ {[}Fig.~\foreignlanguage{english}{\ref{cap:DFT-electrode-construction}}(c)].
Details on the symmetrization procedure and the transformation of
crystal parameters under rotations are presented in App.~\ref{sec:electrodeDFT-electrode-description}.
The parameters $S_{j0}^{\left(X\right)}$, $H_{j0}^{\left(X\right)}$
are finally employed to construct the semi-infinite crystals and to
obtain the surface Green's function $g_{XX}^{r}$ {[}Fig.~\foreignlanguage{english}{\ref{cap:DFT-electrode-construction}}(d)].
Due to the finite range of the couplings $S_{CX}$, $H_{CX}$, we
need to determine $g_{XX}^{r}$ for the first few surface layers only
{[}blue-shaded regions in Fig.~\ref{cap:QTscheme}(a)]. We compute
these with the help of the decimation technique of Ref.~\onlinecite{Guinea:PRB1983},
which we have generalized to deal with the nonorthogonal basis sets.\cite{Pauly:Thesis2007}
The parameters $S_{j0}$, $H_{j0}$ can be computed once for a given
metal and can then be used in transport calculations with electrodes
of various spatial orientations.%

For Au ($a_{0}=4.08$ \AA) we have analyzed spheres ranging between
13 and 429 atoms, while for Al ($a_{0}=4.05$ \AA) they vary between
13 and 555 atoms. Since we want to describe bulk, the parameters extracted
from the largest clusters will obviously provide the best description.
There is, however, an additional criterion, which necessitates the
use of large metal clusters for a reliable description of the electrodes.
As discussed in App.~\ref{sub:electrodeDFT-SizeRequire}, it is based
on the positive-semidefiniteness of the bulk overlap matrix. We find
a strong violation of this criterion, if the extraction of parameters
is performed such that only the couplings of the cental atom to its
nearest neighbors are considered. As a further demonstration of the
quality of our description, we show in App.~\ref{sub:DFT-DOS} the
convergence of the DOS with respect to $R^{sphere}$.

For the transport calculations we need a value for the Fermi energy.
The biggest Au and Al spheres computed, Au$_{429}$ and Al$_{555}$,
respectively, are very metallic. They exhibit differences between
the highest occupied molecular orbital (HOMO) and the lowest unoccupied
molecular orbital (LUMO) of less than $0.06$ eV. Therefore we set
$E_{F}$ midway between these energies. In this way we obtain $E_{F}=-5.0$
eV for Au and $E_{F}=-4.3$ eV for Al. The values will be used in
all the results below. Notice that the negative values of $E_{F}$
agree well with experimental work functions of $5.31$ to $5.47$
eV for Au and $4.06$ to $4.26$ eV for Al.\cite{Lide:CRC1998}

\section{Metallic atomic contacts\label{sec:AuAlcontacts}}

In this chapter we explore the conduction properties of metallic atomic
contacts of Au and Al. These systems, in particular atomic-sized Au
contacts, have been studied in detail both experimentally and theoretically,
and can therefore be used to test our method. We start by discussing
the transport properties of the Au contacts, consisting of a four-atom
chain, a three-atom chain, and a two-atom chain or {}``dimer''.
Since Al does not form such chains, we consider only a single-atom
contact. For all systems we analyze the transmission, its channel
decomposition, and, in order to obtain knowledge about the conduction
mechanism, the LDOS for atoms in the narrowest part of the contact.
Moreover, we demonstrate the robustness of our transmission curves
with respect to different partitionings of the large ECCs.

\selectlanguage{english}

\subsection{Gold contacts\label{sub:Aucontacts}}

\begin{figure}
\selectlanguage{english}
\begin{centering}\includegraphics[width=0.8\columnwidth,clip=]{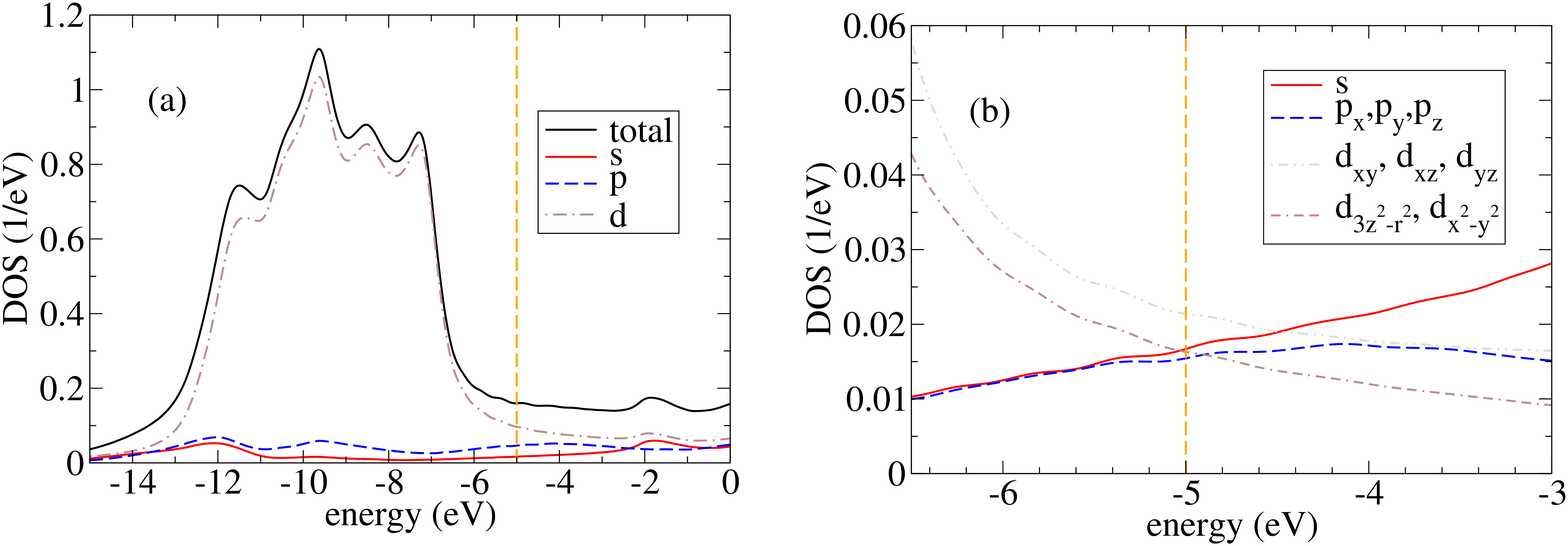}\par\end{centering}

\selectlanguage{american}

\caption{\label{cap:Au-DOSAu}DOS for Au. (a) DOS resolved into $s$, $p$,
and $d$ contributions and (b) DOS resolved into all individual orbital
components. The dashed vertical line indicates $E_{F}=-5.0$ \foreignlanguage{english}{eV}.}
\end{figure}

\selectlanguage{american}
Let us first discuss the electronic structure of Au, where we display
the DOS in Fig.~\ref{cap:Au-DOSAu}.%
 The Fermi energy at $E_{F}=-5.0$ eV is located in a fairly structureless,
flat region somewhat above the $d$ band. Based on the electronic
configuration {[}Xe].$4f^{14}.5d^{10}.6s^{1}$ of the atom, one might
have expected a strong contribution only from the $s$ orbitals at
$E_{F}$. But, as is visible from Fig.~\ref{cap:Au-DOSAu}(b), $s$,
$p$, and $d$ states all yield comparable contributions.%
\footnote{The orbital contributions are obtained by summing over all the basis
functions of a certain angular symmetry: $p_{x}$, for example, results
from a sum over all the $p_{x}$ functions of the basis set, and $p$
is the sum over the $p_{x}$, $p_{y}$, and $p_{z}$ components. %
} This signifies that valence orbitals hybridize strongly in the metal.

When an atomic contact of Au is in a dimer or atomic chain configuration,
a conductance of around $1G_{0}$ is expected from experimental measurements\cite{Scheer:Nature1998,Ludoph:PRB2000,AIYansonPhD:Leiden2001,Yanson:PRL2005,Thijssen:PRL2006}
as well as from theoretical studies.\cite{Cuevas:PRL1998b,Brandbyge:PRB2002,Mozos:Nanotec2002,Lee:PRB2004,Dreher:PRB2005}
The analysis shows that this value of the conductance is due to a
single, almost fully transparent transmission channel. It arises dominantly
from the $s$ orbitals of the noble metal Au, since the electronic
structure in the narrowest part of the contact resembles more the
electronic configuration of the atom.\cite{Scheer:Nature1998,Cuevas:PRL1998b}

\subsubsection{Determination of contact geometries}

\begin{figure}[b]
\begin{centering}\includegraphics[width=0.5\columnwidth,clip=]{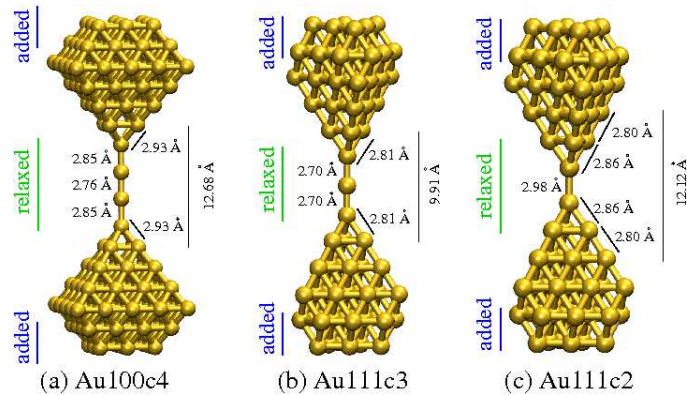}\par\end{centering}

\caption{\label{cap:Au-geosAu}ECCs for Au. (a) Au100c4 is a four-atom Au
chain, (b) Au111c3 is a three-atom Au chain, and (c) Au111c2 is a
two-atom Au chain or dimer contact. For (a) electrodes are oriented
in the {[}$100$] direction, while for (b) and (c) this is the {[}$111$]
direction. The main crystallographic direction is assumed to be parallel
to the $z$ axis. Indicated are also the most important bond distances
together with information on the construction of the ECCs.}
\end{figure}

Despite the consensus that the conductance of atomic chains of Au
is around $1G_{0}$, the precise atomic positions play an important
role.\cite{Yanson:Nature1998,Dreher:PRB2005} Therefore it is necessary
to construct reference geometries that have been studied with a well-established
transport method. We choose to compare to results obtained with TRANSIESTA.\cite{Brandbyge:PRB2002}
The ECCs investigated are shown in Fig.~\ref{cap:Au-geosAu}.%
 The four-atom Au chain with electrodes oriented in the {[}$100$]
direction, called Au100c4, corresponds to a contact geometry examined
in Ref.~\onlinecite{Frederiksen:PRL2004} {[}see Fig.~1(b) therein].
The three-atom Au chain, Au111c3, is similar to a configuration in
Fig.~9(d) of Ref.~\onlinecite{Brandbyge:PRB2002}. In addition,
we study a Au dimer contact, Au111c2, where a two-atom chain is forming
the narrowest part. In contrast to Au100c4, for the latter two contacts
the electrodes are along the {[}$111$] direction.

Let us briefly explain, how we determine these geometries (Fig.~\ref{cap:Au-geosAu}).
For Au100c4 we construct two ideal, atomically sharp Au {[}$100$]
pyramids, with two atoms in between. The pyramids end with the layer
consisting of 25 atoms. The distance between the layers containing
four atoms is set to $12.68$ \AA\ {[}Fig.~\ref{cap:Au-geosAu}(a)],
as in Ref.~\onlinecite{Frederiksen:PRL2004}. Next we relax the four
chain atoms without imposing symmetries, keeping all other atoms fixed.
After geometry optimization, we find that the configuration agrees
well with symmetry $D_{4h}$. We add two more Au layers with 16 and
9 atoms on each side, where the ECC now consists of 162 atoms, and
perform a final DFT calculation, exploiting the symmetry $D_{4h}$.
As compared to Ref.~\onlinecite{Frederiksen:PRL2004}, all bond distances
indicated in Fig.~\ref{cap:Au-geosAu}(a) agree to within $0.01$
\AA, except for the distance between the central chain atoms, where
our distance is shorter by $0.07$ \AA. For Au111c3 we proceed similarly
to Au100c4 {[}Fig.~\ref{cap:Au-geosAu}(b)]. We start with two perfect
Au {[}$111$] pyramids, set the distance between the Au layers with
3 atoms to $9.91$ \AA,\cite{Brandbyge:PRB2002} and cut the pyramid
off at the layers containing $10$ atoms. Then we add one atom in
the middle, relax the three chain atoms, add two layers on each side
with 12 and 6 atoms, and perform a calculation in symmetry $D_{3d}$.
Au111c3 consists of 77 atoms in total. Our bond distances agree with
those in Fig.~9(d) of Ref.~\onlinecite{Brandbyge:PRB2002} to within
$0.02$ \AA. For Au111c2 we include also the first Au layer in the
geometry optimization process. The distance between the fixed layers
with 6 atoms is $12.12$ \AA. Otherwise the steps are the same as
for Au111c3. The ECC is computed in symmetry $D_{3d}$ and consists
of 76 atoms. In the parts excluded from the geometry optimization,
atoms are all positioned on the bulk fcc lattice, where we set the
lattice constant to the experimental value of $4.08$ \AA, which
corresponds to a nearest-neighbor distance of $2.88$ \AA. In each
ECC the main crystallographic direction is aligned with the $z$ axis,
which is the transport direction.

\subsubsection{Four-atom gold chain}

\begin{figure}
\selectlanguage{english}
\begin{centering}\includegraphics[width=0.8\columnwidth,clip=]{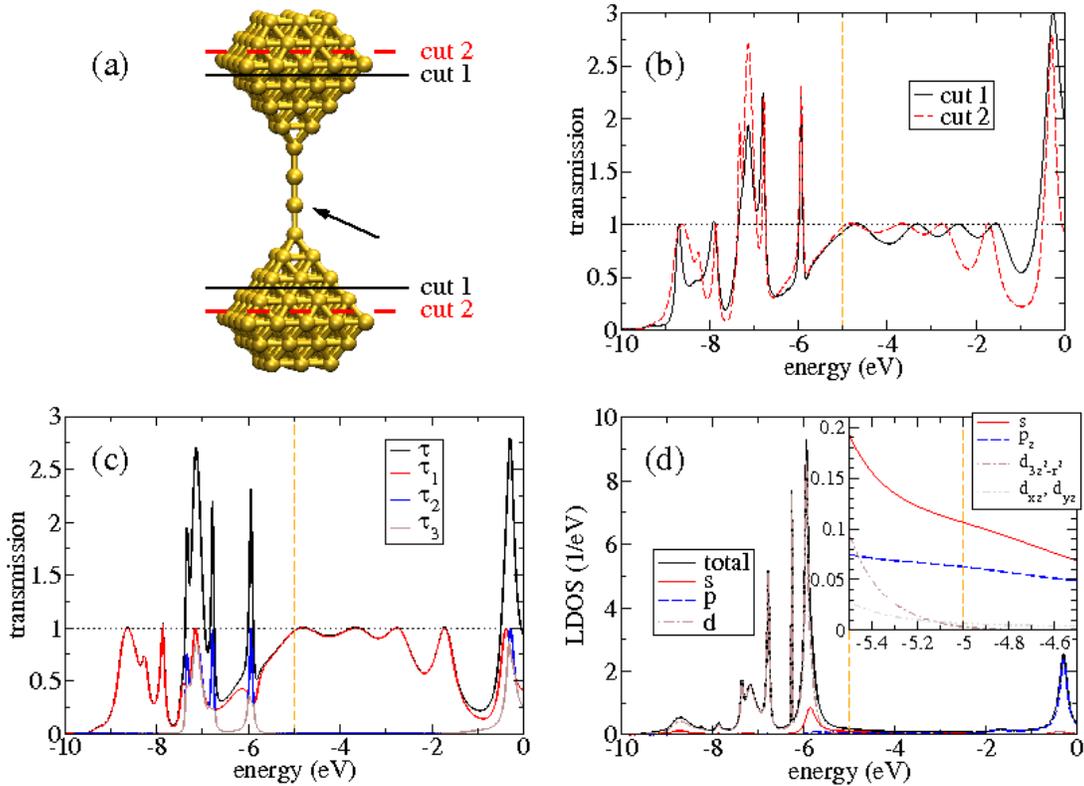}\par\end{centering}

\selectlanguage{american}

\caption{\label{cap:Au-Au100c4-conf-cuts-chan-LDOS}Au100c4. (a) ECC with
two different partitionings into $L$, $C$, and $R$ regions, cuts
1 and 2, and (b) the transmission as a function of the energy for
these cuts. For cut 2 (c) transmission resolved into its transmission
channels and (d) LDOS of the chain atom indicated in (a).}
\end{figure}

Let us now study the conduction properties for the contact Au100c4
{[}Fig.~\ref{cap:Au-Au100c4-conf-cuts-chan-LDOS}(a)].%
 There are different possibilities to partition the ECC into the $L$,
$C$, and $R$ regions. The cuts should be done so that $L$ and $R$
are unconnected ($S_{LR}=0$ and $H_{LR}=0$, see Sec.~\ref{sub:Transport-formalism}).
Hence the $C$ region must be long enough. In order to describe well
the coupling to the electrode surface (Fig.~\ref{cap:QTscheme}),
it is furthermore necessary to have sufficiently many layers in the
$L$ and $R$ regions. We observe that at least two layers are needed
to obtain reasonable transmission curves. For the two different cuts
of Fig.~\ref{cap:Au-Au100c4-conf-cuts-chan-LDOS}(a) $\tau(E)$ is
plotted in Fig.~\ref{cap:Au-Au100c4-conf-cuts-chan-LDOS}(b). In
both cases it is found to be almost identical, indicating a sufficient
robustness of our method. The transmissions at the Fermi energy are
$\tau(E_{F})=0.93$ and $0.98$ for cuts 1 and 2, respectively. These
values correspond well to the result $\tau(E_{F})=0.99$ of Ref.~\onlinecite{Frederiksen:PRL2004}. 

For cut 2 it is visible in Fig.~\ref{cap:Au-Au100c4-conf-cuts-chan-LDOS}(c)
that the transmission at $E_{F}$ is dominated by a single channel,
in good agreement with experimental observations\cite{Scheer:Nature1998}
and previous theoretical studies.\cite{Brandbyge:PRB2002,Mozos:Nanotec2002}
In general, the electronic structure at the narrowest part should
have the most decisive influence on the conductance of an atomic contact.
Therefore we plot in Fig.~\ref{cap:Au-Au100c4-conf-cuts-chan-LDOS}(d)
the LDOS of the atom indicated by the arrow in Fig.~\ref{cap:Au-Au100c4-conf-cuts-chan-LDOS}(a),
resolved in its individual orbital contributions. Compared to the
bulk DOS of Fig.~\ref{cap:Au-DOSAu}, it is dominated by $s$ at
$E_{F}$, where the contributions of all other orbitals than $s$
and $p_{z}$ are suppressed. These two orbitals will form the almost
fully transparent transmission channel, which is radially symmetric
with respect to the $z$ axis.

\subsubsection{Three-atom gold chain}

\begin{figure}
\selectlanguage{english}
\begin{centering}\includegraphics[width=0.8\columnwidth,clip=]{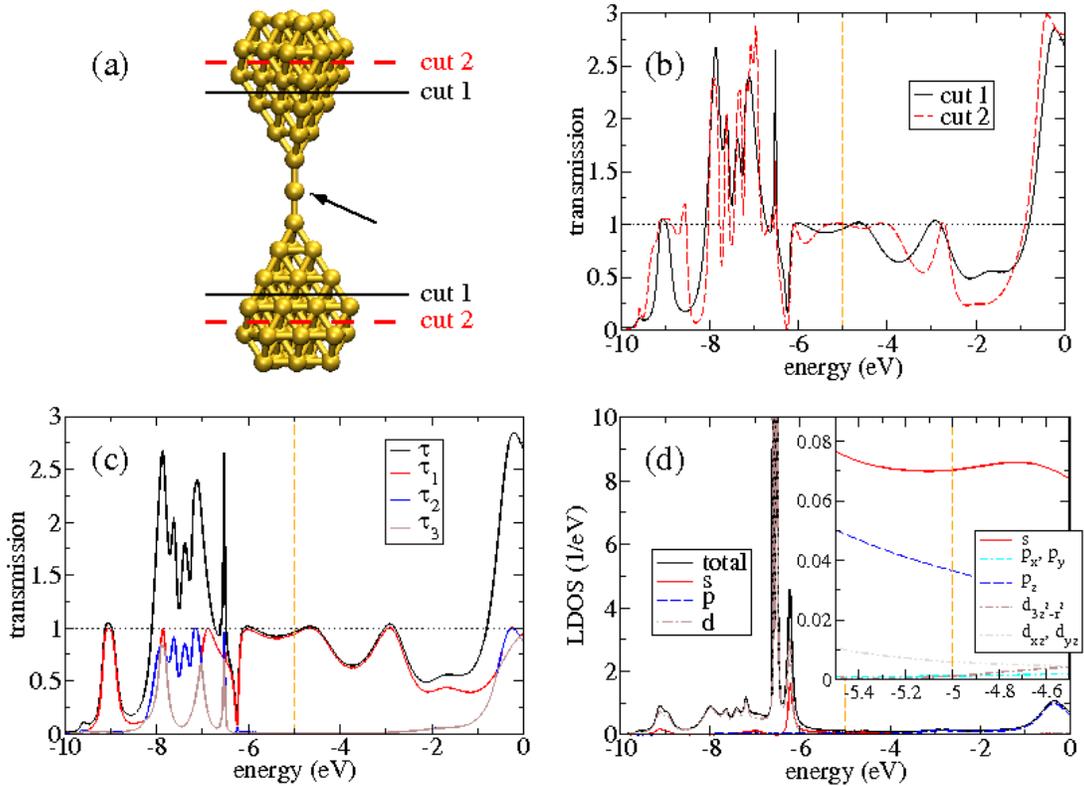}\par\end{centering}

\selectlanguage{american}

\caption{\label{cap:Au-Au111c3-conf-cuts-chan-LDOS}Au111c3. (a) ECC with
two different partitionings into $L$, $C$, and $R$ regions, cuts
1 and 2, and (b) the transmission as a function of the energy for
these cuts. For cut 1 (c) transmission resolved into its transmission
channels and (d) LDOS of the central chain atom indicated in the ECC.}
\end{figure}

Exactly the same analysis will now be carried out for the contact
Au111c3. In Fig.~\ref{cap:Au-Au111c3-conf-cuts-chan-LDOS} the geometry
of the ECC, the transmission for different partitionings, the transmission
channels, and the LDOS of the central chain atom are shown.%
 As for Au100c4, we observe that the different cuts yield very similar
transmission curves {[}Fig.~\ref{cap:Au-Au111c3-conf-cuts-chan-LDOS}(b)].
Furthermore all the basic features in $\tau(E)$ are the same as in
the TRANSIESTA calculation {[}see Fig.~11(d) of Ref.~\onlinecite{Brandbyge:PRB2002}].
Above the $d$ band, which exhibits a very narrow and high final peak,
there is a dip in $\tau(E)$ in both cases. The transmission recovers,
however, and a flat region with a value of around one is visible.
At the Fermi energy cuts 1 and 2 yield $\tau(E_{F})=0.96$ and $0.99$,
respectively. This is in reasonable agreement with $\tau(E_{F})=0.94$
in Ref.~\onlinecite{Brandbyge:PRB2002}, considering the differences
in the electrode geometry, basis set, and exchange-correlation functional.
We observe from Fig.~\ref{cap:Au-Au111c3-conf-cuts-chan-LDOS}(c)
for cut 1 that the transmission at $E_{F}$ is dominated by a single
transmission channel, and the LDOS indicates a dominant contribution
of $s$ orbitals {[}Fig.~\ref{cap:Au-Au111c3-conf-cuts-chan-LDOS}(d)].
In addition, the peak structures in $\tau(E)$ for the $d$ states
correspond well to peaks in the LDOS. This observation can also be
made in Figs.~\ref{cap:Au-Au100c4-conf-cuts-chan-LDOS}(c) and \ref{cap:Au-Au100c4-conf-cuts-chan-LDOS}(d)
for Au100c4.

\subsubsection{Two-atom gold chain}

\begin{figure}
\selectlanguage{english}
\begin{centering}\includegraphics[width=0.9\columnwidth,clip=]{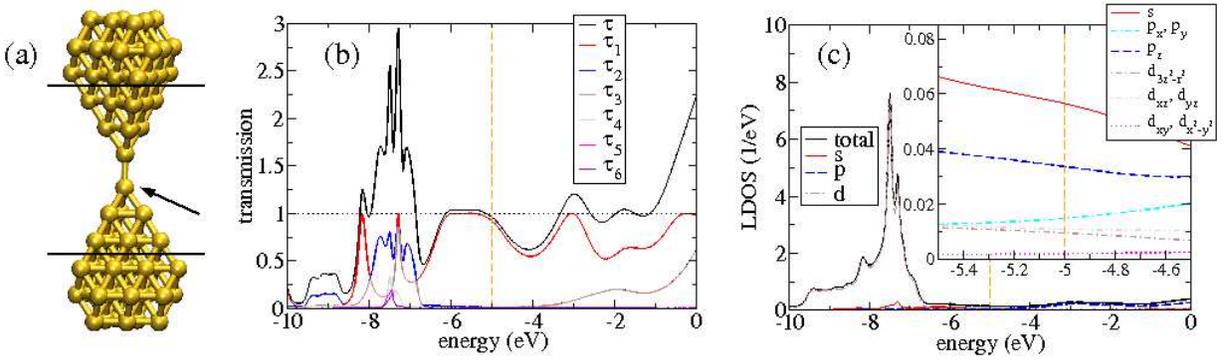}\par\end{centering}

\selectlanguage{american}

\caption{\label{cap:Au-Au111c2-conf-chan-LDOS}Au111c2. (a) ECC, and for the
cut and atom indicated (b) the transmission resolved into its transmission
channels and (c) the LDOS with its individual orbital contributions,
respectively.}
\end{figure}

The transmission and LDOS resolved into transmission channels and
orbital components, respectively, are shown in Fig.~\ref{cap:Au-Au111c2-conf-chan-LDOS}
for the dimer contact Au111c2.%
 As for Au100c4 and Au111c3 we observe a single dominant channel at
$E_{F}$, and $\tau(E_{F})=0.96$. The finding of such a dominant
channel for chains of two or more atoms is in good agreement with
our analysis of less symmetric contacts, which were based on a combination
of a tight-binding model and classical molecular dynamics simulations.\cite{Dreher:PRB2005}
However, that $\tau(E)$ increases partly even above one in the vicinity
of $E_{F}$, signals that the influence of other channels is increased
as compared to Au100c4 and Au111c3. Indeed, the LDOS of the atom in
the narrowest part of the constriction {[}Figs.~\ref{cap:Au-Au111c2-conf-chan-LDOS}(a)
and \ref{cap:Au-Au111c2-conf-chan-LDOS}(c)] shows in particular increased
$p_{x}$ and $p_{y}$ contributions. Also, the $d$ states exhibit
a less pronounced peak structure than was visible in Figs.~\ref{cap:Au-Au100c4-conf-cuts-chan-LDOS}(d)
and \ref{cap:Au-Au111c3-conf-cuts-chan-LDOS}(d). This is due to the
higher coordination number of the atom and the enhanced coupling to
the electrodes. 

\selectlanguage{english}

\subsection{Aluminum contacts\label{sub:Alcontacts}}

\begin{figure}[b]
\selectlanguage{english}
\begin{centering}\includegraphics[width=0.8\columnwidth,clip=]{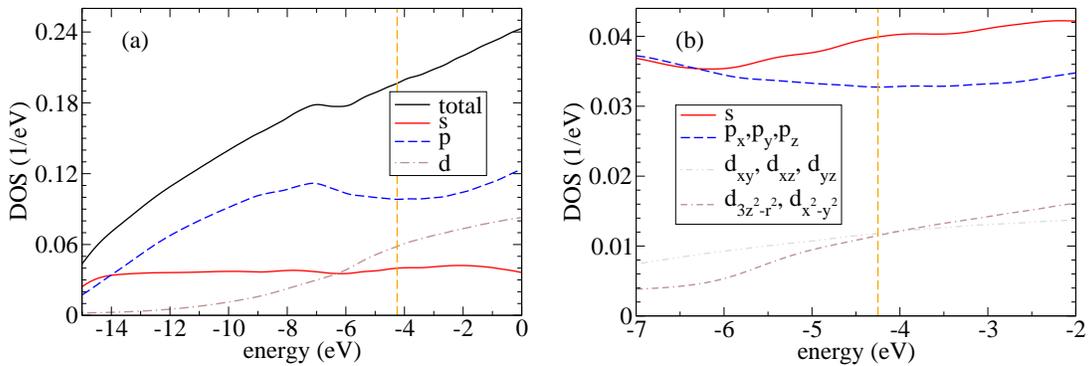}\par\end{centering}

\selectlanguage{american}

\caption{\label{cap:Al-DOSAl}DOS for Al. (a) DOS resolved into $s$, $p$,
and $d$ contributions and (b) DOS resolved into all individual orbital
components. The dashed vertical line indicates $E_{F}=-4.3$ eV.}
\end{figure}

\selectlanguage{american}
As is visible from the bulk DOS in Fig.~\ref{cap:Al-DOSAl}, the
electronic structure of Al differs substantially from that for Au.%
 While the latter is a noble metal with an $s$ valence, the Al atom
has the electronic configuration {[}Ne].$3s^{2}.3p^{1}$ with an open
$p$ shell, and the metal is hence considered $sp$-valent. The strong
contribution of $s$ and $p$ states is also observed in the DOS,
where $d$ states play only a minor role. As compared to Au, the DOS
exhibits a noticeable energy dependence around $E_{F}$.

For Al we study an ideal fcc {[}$111$] pyramid, consisting of 251
atoms {[}Fig.~\ref{cap:Al-Al111c1-conf-chan-LDOS}(a)], henceforth
referred to as Al111c1. Ideal means that the atoms are positioned
on an fcc lattice with the experimental lattice constant $a_{0}=4.05$
\AA. We have already reported results for Al dimer contacts in Ref.~\onlinecite{Wohlthat:PRB2007}.

\subsubsection{Aluminum single-atom contact}

\begin{figure}
\selectlanguage{english}
\begin{centering}\includegraphics[width=0.8\columnwidth,clip=]{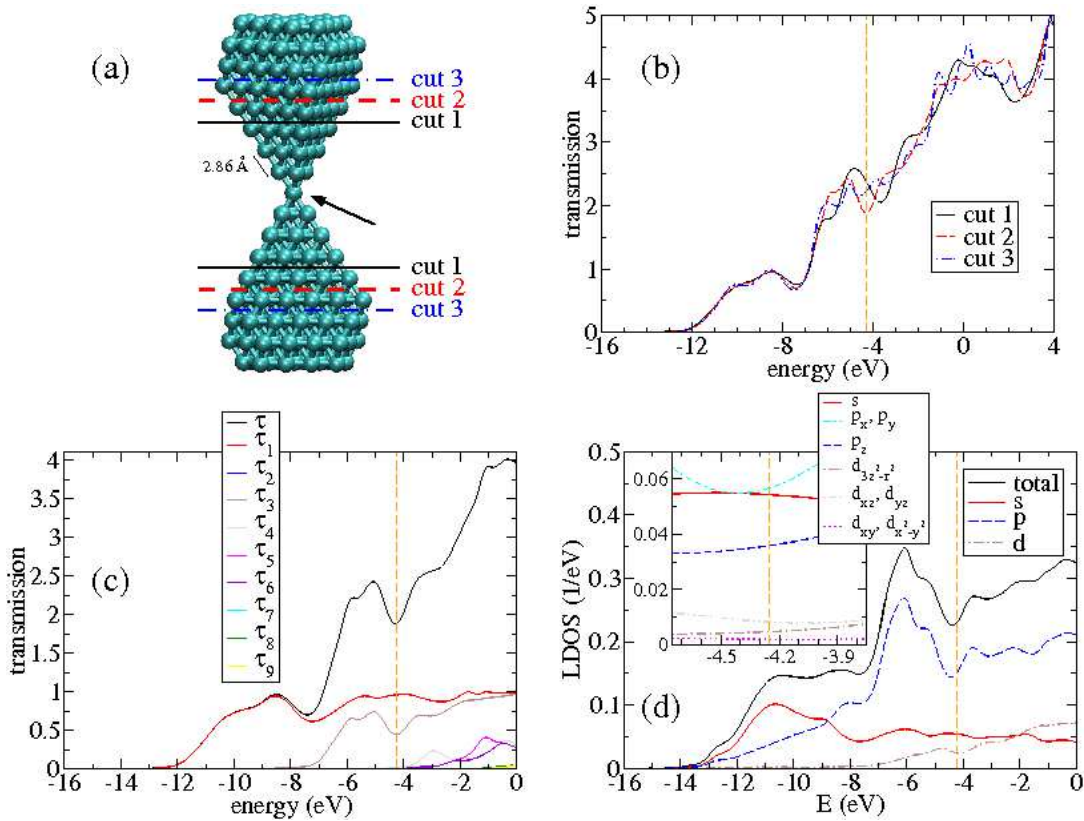}\par\end{centering}

\selectlanguage{american}

\caption{\label{cap:Al-Al111c1-conf-chan-LDOS}Al111c1. (a) ECC with three
different partitionings into $L$, $C$, and $R$ regions, cuts 1,
2, and 3, and (b) the transmission as a function of the energy for
these cuts. For cut 2 (c) the transmission resolved into its transmission
channels and (d) the LDOS of the central atom, indicated by the arrow
in (a). The nearest-neighbor distance shown in the ECC is identical
for all atoms.}
\end{figure}

In Fig.~\ref{cap:Al-Al111c1-conf-chan-LDOS} the transmission is
displayed for three different partitionings of the ECC Al111c1.%
 Also shown are the transmission channels and the LDOS of the atom
in the narrowest part of the contact for a selected cut. For energies
below $-6$ eV, there are practically no differences visible between
the curves for the three different partitionings. Nevertheless, some
deviations arise at $E_{F}$, and we obtain $\tau(E_{F})=2.36$ (cut
1), $1.88$ (cut 2), and $2.23$ (cut 3). Similar to Au, we attribute
these 20\% relative variations to spurious scattering at the $LC$
and $CR$ interfaces. Our values for $\tau(E_{F})$ of around two
agrees nicely with those reported for single-atom contacts in Ref.~\onlinecite{Jelinek:PRB2003}.
Compared to Au, the transmission-channel structure has changed in
an obvious way. There are three channels at $E_{F}$, which is in
line with experimental observations of Refs.~\onlinecite{Scheer:PRL1997,Scheer:Nature1998}.
Due to the $D_{3d}$ symmetry of the ECC, transmission-channel degeneracies
arise, where in particular $\tau_{2}=\tau_{3}$. As is visible from
the LDOS, these additional channel contributions mainly stem from
the $p_{x}$ and $p_{y}$ orbitals, while $s$ and $p_{z}$ are forming
the nondegenerate $\tau_{1}$.\cite{Scheer:Nature1998,Cuevas:PRL1998b,Pauly:PRB2006}

\selectlanguage{english}

\section{Conclusions\label{sec:Conclusions}}

We have developed a cluster-based method to study the charge transport
properties of molecular and atomic contacts. We treat the electronic
structure at the level of DFT, and describe transport in terms of
the Landauer formalism expressed with standard Green's function techniques.
Special emphasis is placed on the modeling of the electrodes and the
construction of the associated bulk parameters from spherical metal
clusters. We showed that these clusters need to be sufficiently large
to produce reliable bulk parameters, where a criterion for the extent
of the spherical clusters is set by the overlap of the nonorthogonal
basis functions. In our studies we crucially rely on the accurate
and efficient quantum-chemical treatment of systems consisting of
several hundred atoms, made possible by use of the quantum chemistry
package TURBOMOLE. Compared to supercell approached, our method has
the advantage that we genuinely describe single-atom or single-molecule
contacts.

\selectlanguage{american}
As an application of our method we analyzed Au and Al atomic contacts.
Studying a four-, a three-, and a two-atom chain with varied electrode
lattice orientations for Au, we found a conductance close to $1G_{0}$,
carried by a single transmission channel. Next we investigated an
ideal Al single-atom contact, and found three transmission channels
to contribute significantly to the conductance of around $2G_{0}$.
These results are in good agreement with previous experimental and
theoretical investigations. Both for Au and Al we demonstrated the
robustness of our transmission curves with respect to partitionings
of the contact systems. The results illustrate the applicability of
our method to various electrode materials.

Beside the metallic atomic contacts examined here, the presented method
has been applied in the field of molecular electronics. Studies include
the dc conduction properties of dithiolated-oligophenylene and diamino-alkane
junctions\cite{Viljas:PRB2007b,Pauly:PRB2008,Pauly:arXiv:0709,Wohlthat:ChemPhysLett2008}
as well as oxygen adsorbates in Al contacts.\cite{Wohlthat:PRB2007}
In addition, the thermopower\cite{Pauly:arXiv:0709} and photoconductance\cite{Viljas:PRB2007b}
of molecular junctions has been investigated in this way. Our studies
demonstrate the value of parameter-free modeling for understanding
transport at the molecular and atomic scale.

\begin{acknowledgments}
We thank R.~Ahlrichs for providing us with TURBOMOLE and acknowledge
stimulating discussions with him and members of his group, in particular
N.~Crawford, F.~Furche, M.~Kattannek, P.~Nava, D.~Rappoport,
C.~Schrodt, M.~Sierka, and F.~Weigend. This work was supported
by the Helmholtz Gemeinschaft (Contract No.~VH-NG-029), by the EU
network BIMORE (Grant No.~MRTN-CT-2006-035859), and by the DFG within
the CFN and SPP 1243. M.~H.~acknowledges funding by the Karlsruhe
House of Young Scientists and F.~P.~that of a Young Investigator
Group at KIT.
\end{acknowledgments}
\selectlanguage{english}
\appendix

\section{Nonorthogonal basis sets\label{sec:NO-basis-sets}}

For practical reasons one often employs nonorthogonal basis sets in
quantum-chemical calculations, consisting for example of a finite
set of Gaussian functions. The electronic structure is described in
the spirit of the linear combination of atomic orbitals (LCAO),\cite{Szabo:Dover1996,Koch:Wiley2001,Atkins:Oxford1999}
and this is also how TURBOMOLE is implemented. While it is in principle
always possible to transform to an orthogonal basis, it may be more
convenient to work directly with the nonorthogonal states.

A concise mathematical description using nonorthogonal basis states
can be formulated in terms of tensors. The formalism is presented
in a fairly general form in Ref.~\onlinecite{Artacho:PRA1991}, where
also the modifications of second quantization are addressed. Below
we discuss some of the subtleties related to the use of nonorthogonal
basis functions that are important for our method.\cite{Pauly:Thesis2007}
Since the basis functions are real-valued in our case, the full complexity
of the tensor formalism is not needed.\cite{Iben:Teubner1999,Borisenko:Dover1979}
Furthermore we use a simplified notation, where all tensor indices
appear as subscripts of matrices.

\subsection{Current formula for nonorthogonal, local basis sets}

The most important quantity for transport calculations is the electric
current. In the NEGF formalism, its determination requires a separation
of the contact into subsystems similar to Fig.~\ref{cap:QTscheme}(a).\cite{Caroli:SSP1971,Meir:PRL1992,Datta:Cambridge2005}
However, due to the overlap of the basis functions in a nonorthogonal
basis, the charges of the subsystems are not well defined. Different
ways of determining them exist, e.g.~the Mulliken or L\"owdin population
analysis.\cite{Szabo:Dover1996} Despite these additional complications,
the Landauer formula {[}Eq.~(\ref{eq:conductance_Tfinite})] can
be derived in a similar fashion as for an orthogonal basis. Recent
discussions of the derivation can be found in Refs.~\onlinecite{Viljas:PRB2005,Thygesen:PRB2006}.

\subsection{Single-particle Green's functions\label{sec:SingleParticleG}}

Consider the single-particle Hamiltonian $H$ describing the entire
system. The retarded Green's operator is defined as $G^{r}(E)=\left[\left(E+i0^{+}\right)\mathbbm{1}-H\right]^{-1}$.
Now consider the local, nonorthogonal basis $|i\rangle$ with the
(covariant) matrix elements of the overlap $S_{ij}=\left\langle i\right.\left|j\right\rangle $
and the Hamiltonian $H_{ij}=\left\langle i\right|H\left|j\right\rangle $.\cite{Iben:Teubner1999,Borisenko:Dover1979}
Compared to Sec.~\ref{sub:Transport-formalism} the index $i$, used
throughout this appendix, is a collective index, denoting both the
position at which the basis state is centered and its type. The components
of the retarded Green's function, defined by $G^{r}=\sum_{i,j}|i\rangle G_{ij}^{r}\langle j|$,%
\footnote{\selectlanguage{english}
Note that the $G_{ij}^{r}$ are the contravariant components of $G^{r}$.\cite{Lohez:PRB1983,Sulston:PRB2003,Iben:Teubner1999,Borisenko:Dover1979}
In the conventional tensor formulation they would be written $\left(G^{r}\right)^{ij}$,
but in our simplified notation no distinction between co- and contravariant
components is being made.\selectlanguage{american}
} satisfy the equation\cite{Xue:ChemPhys2002} \begin{equation}
\sum_{k}\left[\left(E+i0^{+}\right)S_{jk}-H_{jk}\right]G_{kl}^{r}(E)=\delta_{jl}.\label{eq:ES-HG}\end{equation}

The Green's function $G_{CC}^{r}$ is defined as $G_{ij}^{r}$ restricted
to the central region $C$. It can be calculated according to Eq.~(\ref{eq:GrCC}).
Due to the nonorthogonal basis, the perturbation that couples $C$
to the lead $X=L,R$ and enters the self-energy {[}Eq.~(\ref{eq:Simga_X})],
is given by $H_{CX}-ES_{CX}$ and thus includes also an overlap contribution.
It is interesting to observe that as $E\rightarrow\infty$ the self-energies
and the Green's function behave as\begin{eqnarray}
\Sigma_{X}^{r}(E) & \underrightarrow{E\rightarrow\infty} & ES_{CX}\left(S_{XX}\right)^{-1}S_{XC}\label{eq:SigmaX_Easymptotic}\\
G_{CC} & \underrightarrow{E\rightarrow\infty} & E^{-1}\left(S^{-1}\right)_{CC}\label{eq:GCC_Easymptotic}\end{eqnarray}
 with \begin{equation}
\left(S^{-1}\right)_{CC}=\left[S_{CC}+\sum_{X=L,R}S_{CX}\left(S_{XX}\right)^{-1}S_{XC}\right]^{-1}.\label{eq:S_inverse}\end{equation}
 Thus also the inverse overlap matrix of $C$ is {}``renormalized''
due to the coupling to the leads.

\subsection{Local density of states\label{sub:noG-LDOS}}

Using a set of orthonormal energy eigenstates $\left|\mu\right\rangle $
that satisfy $H\left|\mu\right\rangle =\varepsilon_{\mu}\left|\mu\right\rangle $,
we obtain the decomposition $\rho_{\mu\nu}(E)=\langle\mu|\rho(E)|\nu\rangle=\sum_{\mu}\delta(E-\varepsilon_{\mu})\delta_{\mu\nu}$
of the spectral density defined by Eq.~(\ref{eq:rhoEdef}). Clearly
$\rho_{\mu\nu}(E)$ fulfills the normalization \begin{equation}
\int_{-\infty}^{\infty}dE\rho_{\mu\nu}(E)=\delta_{\mu\nu}.\label{eq:intRho_mu_nu}\end{equation}
If, instead, we consider the components defined by $\rho_{ij}(E)=-\mathrm{Im}\left[G_{ij}^{r}(E)\right]/\pi$,
where $G_{ij}^{r}(E)$ is given by Eq.~(\ref{eq:ES-HG}), we find
\begin{equation}
\int_{-\infty}^{\infty}dE\rho_{ij}(E)=\left(S^{-1}\right)_{ij}.\label{eq:intImGij_contra}\end{equation}
The normalization of Eq.~(\ref{eq:intRho_mu_nu}) can be recovered
by performing a L\"owdin orthogonalization of the basis \begin{equation}
\int_{-\infty}^{\infty}dE(S^{1/2})_{ik}\rho_{kl}(E)(S^{1/2})_{lj}=\delta_{ij}\label{eq:intRho_i_j_Shalf}\end{equation}

Let us analyze the LDOS of the central region $\rho_{CC}(E)=-\mathrm{Im}\left[G_{CC}^{r}(E)\right]/\pi$,
which is $\rho_{ij}(E)$ restricted to $C$. Analogously to Eq.~(\ref{eq:intRho_i_j_Shalf}),
we have defined the LDOS at atom $i$ and its decomposition into orbitals
$\alpha$ in Eqs.~(\ref{eq:LDOS_i}) and (\ref{eq:LDOS_ia}). Since
$\rho_{CC}(E)$ is a positive-semidefinite matrix, it is easy to show
that $\mathrm{LDOS}_{i\alpha}(E)$ is positive for all $E$. However,
the normalization $\int_{-\infty}^{\infty}dE\mathrm{LDOS}_{i\alpha}(E)=1$
is only approximately fulfilled. This could be corrected by multiplying
in Eq.~(\ref{eq:LDOS_ia}) with $\left(S^{-1}\right)_{CC}^{-1/2}$
{[}Eq.~(\ref{eq:S_inverse})] instead of $S_{CC}^{1/2}$. But since
the self-energy contributions $\sum_{X=L,R}S_{CX}\left(S_{XX}\right)^{-1}S_{XC}$
constitute only a surface correction, their neglect may be justified
for atoms in the middle of $C$.

\section{Description of electrodes\label{sec:electrodeDFT-electrode-description}}

\selectlanguage{american}

\subsection{Size requirement for the cluster construction\label{sub:electrodeDFT-SizeRequire}}

How large do the spherical metal clusters, involved in the construction
in Fig.~\ref{cap:DFT-electrode-construction}, need to be for a convergence
of the bulk parameters? Since the matrix elements of the Hamiltonian
and the overlap decay similarly with increasing interatomic distance,
we can concentrate on the overlap. For it a rather well-defined criterion
can be found: The clusters should so large that the extracted bulk
overlap matrix is positive-semidefinite.

We define states $\left|\vec{k},\alpha\right\rangle =\sum_{j}e^{i\vec{k}\cdot\vec{R}_{j}}\left|j,\alpha\right\rangle $
in $k$-space. Since $S_{i\alpha,j\beta}=\left\langle i,\alpha\right.\left|j,\beta\right\rangle $
is a positive-semidefinite matrix,\cite{Szabo:Dover1996} the same
is true for the overlap in $k$-space \begin{eqnarray}
S_{\alpha\beta}(\vec{k},\vec{k}')=\left\langle \vec{k},\alpha\right.\left|\vec{k}',\beta\right\rangle  & = & \sum_{l,m}e^{-i\vec{k}\cdot\vec{R}_{l}}S_{l\alpha,m\beta}e^{i\vec{k}\cdot\vec{R}_{m}}\label{eq:Skkp_def}\\
 & = & N\delta_{\vec{k},\vec{k}'}S_{\alpha\beta}(\vec{k}),\nonumber \end{eqnarray}
where we used that $S_{l\alpha,m\beta}=S_{\left(l-m\right)\alpha,0\beta}$.
In the expression, $N$ is the number of atoms in the crystal and
\begin{equation}
S_{\alpha\beta}(\vec{k})=\sum_{j}e^{-i\vec{k}\cdot\vec{R}_{j}}S_{j\alpha,0\beta}.\label{eq:Sk_def}\end{equation}

In order to study the positive-semidefiniteness of $S_{\alpha\beta}(\vec{k},\vec{k}')$
it is hence sufficient to investigate the behavior of $S_{\alpha\beta}(\vec{k})$.
To do so for a complex quantum-chemistry basis set, we define the
positive-definiteness measure \begin{equation}
\xi(R^{sphere})=\min_{\vec{k}}(\mathcal{S}(\vec{k})).\label{eq:DFT-xi-def-posdef}\end{equation}
In this expression $\mathcal{S}(\vec{k})$ is the smallest eigenvalue
of the matrix $S_{\alpha\beta}(\vec{k})$, where $S_{\alpha\beta}(\vec{k})$
is constructed from the crystal parameters extracted from a cluster
with radius $R^{sphere}$ {[}$S_{\alpha\beta}(\vec{k})=\sum_{\vec{R}_{j};|\vec{R_{j}}|\leq R^{sphere}}e^{-i\vec{k}\cdot\vec{R}_{j}}S_{j\alpha,0\beta}$].
In the discrete Fourier transformations we assume periodic boundary
conditions with a finite periodicity length along the standard primitive
lattice vectors.\cite{Ashcroft:Harcourt1976,Pauly:Thesis2007} $R^{sphere}$
must be chosen large enough for $\xi$ to be positive or, if $\xi$
remains negative, it must at least be sufficiently small in absolute
value. 

\begin{figure}
\begin{centering}\includegraphics[width=0.8\columnwidth,clip=]{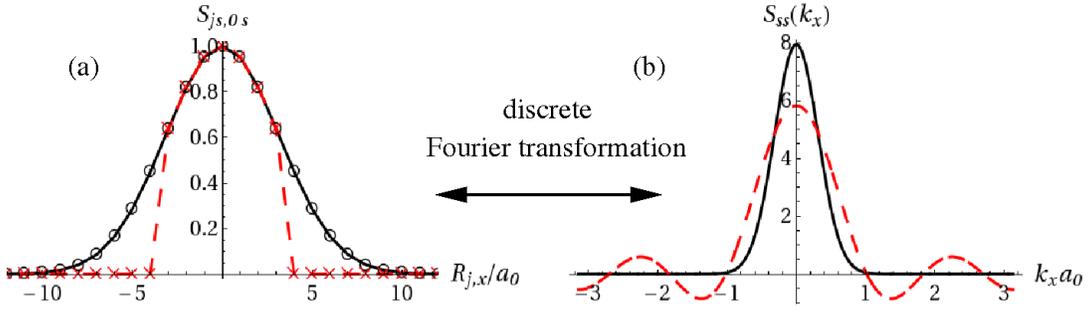}\par\end{centering}

\caption{\label{cap:DFT-s-model-1d-FT}$s$-orbital-chain model. (a) The overlap
$S_{js,0s}$ of an atom with its neighbors at positions $R_{j,x}=j_{1}a_{0}$
with $j_{1}=0,\pm1,\ldots$. (b) Overlap $S_{ss}(k_{x})$ after a
discrete Fourier transformation. In both cases the solid line is for
a large and the dashed line for a small cluster.}
\end{figure}

Let us first illustrate the behavior of $\xi$ at the example of an
$s$-orbital model. Gaussian $s$ functions are described by\begin{equation}
\phi_{s}(\vec{r})=\left(\frac{2\alpha}{\pi}\right)^{3/4}e^{-\alpha\left|\vec{r}\right|^{2}}\label{eq:DFT-s-orbital-GTO}\end{equation}
with an exponent $\alpha$, characterizing the radial decay. Hence,
the overlap between two atoms\begin{equation}
S_{js,0s}=\int d^{3}r\phi_{s}(\vec{r}-\vec{R}_{j})\phi_{s}(\vec{r})=e^{-\alpha R_{j}^{2}/2}\label{eq:DFT-overlap-Sd}\end{equation}
decays with their distance $R_{j}=|\vec{R}_{j}|$ like a Gaussian
function. We consider an infinitely extended chain with atoms at equally
spaced positions along the $x$-axis ($\vec{R}_{j}=j_{1}a_{0}\vec{e}_{x}$).
The overlap from a selected atom to its neighbors drops off exponentially
as shown in Fig.~\ref{cap:DFT-s-model-1d-FT}(a).%
 The Fourier transformation will again result in a Gaussian with purely
positive values $S_{ss}(k_{x})$ {[}Fig.~\ref{cap:DFT-s-model-1d-FT}(b)].
If, however, overlap matrix elements are taken into account only up
to a certain maximum value $|\vec{R}_{j}|\leq R^{sphere}$, as in
a finite cluster, a rough $\sin(k_{x})/k_{x}$-behavior results, where
$S_{ss}(k_{x})$ becomes negative at certain $k$-values. Upon an
increase of $R^{sphere}$, $S_{ss}(k_{x})$ will evolve into a Gaussian
function and $\xi$ will thus approach zero from below. The negative
tails of $S_{ss}(k_{x})$ are unphysical, and our observation implies
that the clusters used to extract bulk parameters (Fig.~\ref{cap:DFT-electrode-construction})
need to be of a sufficiently large radius $R^{sphere}$, in order
to obtain a reliable description of a crystal. Obviously, the magnitude
of $R^{sphere}$ depends on the basis set chosen.

\begin{figure}
\selectlanguage{english}
\begin{centering}\includegraphics[width=0.8\columnwidth,clip=]{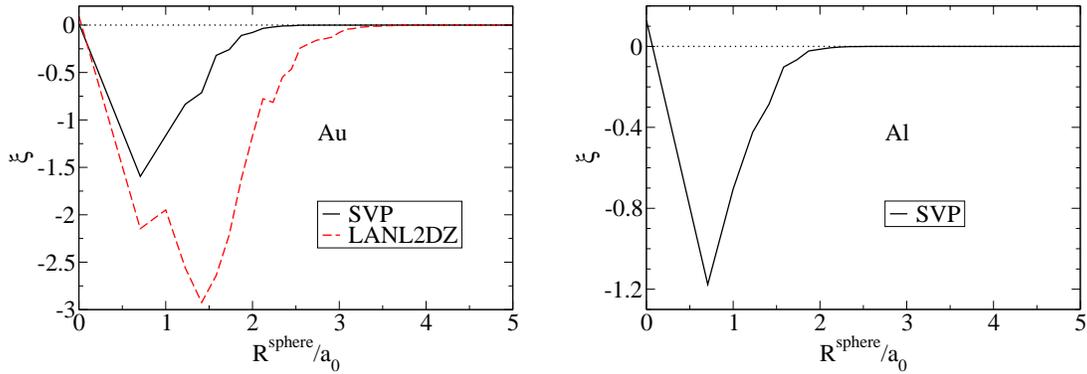}\par\end{centering}

\selectlanguage{american}

\caption{\label{cap:DFT-xi-nshells-Al-Au}The positive-definiteness measure
$\xi$ for Au and Al as a function of $R^{sphere}$. Beside the SVP
basis set, the behavior of $\xi$ is also shown for LANL2DZ, used
in Refs.~\onlinecite{Damle:PRB2001,Damle:ChemPhys2002}. $S(\vec{k})$
{[}Eq.~(\ref{eq:Sk_def})] is evaluated at $32^{3}$ $k$-points.
The radius $R^{sphere}$ has been scaled with the respective lattice
constants ($a_{0}=4.08$ \AA\ for Au and $a_{0}=4.05$ \AA\ for
Al).}
\end{figure}
In Fig.~\ref{cap:DFT-xi-nshells-Al-Au} we plot the behavior of $\xi$
as a function of $R^{sphere}$ for Au and Al. Beside the results for
the SVP basis set, we display $\xi$ for Au also for the basis set
LANL2DZ, used in Refs.~\onlinecite{Damle:PRB2001,Damle:ChemPhys2002}.
It is visible that $\xi$ is positive for a single atom ($R^{sphere}=0$),
but negative for small spheres. With increasing \foreignlanguage{english}{$R^{sphere}$},
$\xi$ approaches $0$ from below similar to the $s$-orbital model.
We find that the elimination of diffuse functions reduces the radius
$R^{sphere}$ for $\xi$ to become positive or negligibly small. For
practical reasons it may happen that $R^{sphere}$ cannot be chosen
large enough to fulfill the positive-semidefiniteness criterion $\xi\geq0$.
In such a case negative eigenvalues of $S_{\alpha\beta}(\vec{k})$
can lead to negative eigenvalues of the scattering-rate matrices $\Gamma_{X}$
{[}Eq.~(\ref{eq:Gamma_X})], since $\rho_{XX}=-\mathrm{Im}\left[g_{XX}^{r}\right]/\pi$
may no longer be positive-semidefinite (see also the discussion in
Sec.~\ref{sub:noG-LDOS}).

\subsection{Bulk densities of states\label{sub:DFT-DOS}}

The DOS can be used as another measure for the convergence to a solid-state
description. With a $k$-space Hamiltonian in an orthogonal basis
set $H^{orth}(\vec{k})$, it is given as \begin{equation}
\mathrm{DOS}(E)=\sum_{\alpha}\mathrm{DOS}_{\alpha}(E)=-\frac{1}{\pi}\mathrm{Tr}_{\alpha}\left[\mathrm{Im}\left[G_{0\alpha,0\alpha}^{orth,r}(E)\right]\right],\label{eq:DFT-DOS_bulk}\end{equation}
where $\alpha$ runs over all basis functions on a bulk atom and $G_{00}^{orth,r}(E)=\int_{BZ}d^{3}kG^{orth,r}(\vec{k},E)/V_{BZ}$
with $G^{orth,r}(\vec{k},E)=\left[E\mathbbm{1}-H^{orth}(\vec{k})\right]^{-1}$
and with the volume $V_{BZ}$ of the first Brillouin zone $BZ$. The
orthogonal Hamiltonian can be obtained in several ways. Two possible
choices are (i) to Fourier transform $S_{j0}$ and $H_{j0}$ and perform
a L\"owdin orthogonalization in $k$-space $H^{orth}(\vec{k})=S^{-1/2}(\vec{k})H(\vec{k})S^{-1/2}(\vec{k})$
or (ii) to construct $H_{j0}^{orth}$, which involves the L\"owdin
transformation $H^{sphere,orth}=\left(S^{-1/2}\right)^{sphere}H^{sphere}\left(S^{-1/2}\right)^{sphere}$
in real space, extraction of $H_{j0}^{sphere,orth}$ and the imposing
of the fcc space group, and to carry out the Fourier transformation
only thereafter. For parameters extracted from large enough clusters
we observe the equivalence of the DOS construction with respect to
the two different orthogonal Hamiltonians $H^{orth}(\vec{k})$. If
$\xi$ remains (slightly) negative due to a too small $R^{sphere}$,
then the construction of the DOS from $H_{j0}^{orth}$ {[}procedure
(ii)] is of a higher quality than that resulting from the L\"owdin
orthogonalization in $k$-space {[}procedure (i)].

\begin{figure}
\selectlanguage{english}
\begin{centering}\includegraphics[width=0.8\columnwidth,clip=]{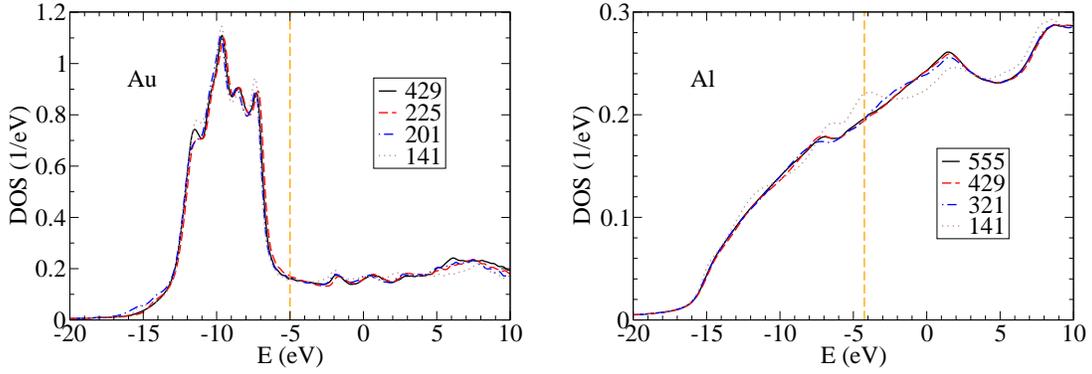}\par\end{centering}

\selectlanguage{american}

\caption{\label{cap:DFT-Bulk-DOS-AuAl}DOS of Au and Al constructed from parameters
extracted from fcc spheres with atom numbers between 141 and 555.
The vertical dashed line indicates the Fermi energy $E_{F}$.}
\end{figure}

In Fig.~\ref{cap:DFT-Bulk-DOS-AuAl} we show the DOS as constructed
via procedure (ii) with parameters extracted from different Au and
Al spheres with 141 to 555 atoms.%
 We observe that the DOS seems well converged with respect to $R^{sphere}$
both for Au and Al for the largest spherical clusters Au$_{429}$
and Al$_{555}$.

\subsection{Transformation of electrode parameters under rotations\label{sec:electrodeDFT-RotTrafo}}

We \foreignlanguage{english}{assume that} two coordinate systems are
connected by the rotation $\varrho$, where $\vec{r}'=\varrho\vec{r}$.
The transformation properties of the electrode parameters \[
Y_{j\alpha,0\beta}=\left\langle j,\alpha\right|Y\left|0,\beta\right\rangle =\int d^{3}r\phi_{\alpha}(\vec{r}-\vec{R}_{j})Y(\vec{r})\phi_{\beta}(\vec{r})\]
\foreignlanguage{english}{with $Y=S,H$ are determined by those of
the basis functions $\left\langle \vec{r}\left.\right|j,\alpha\right\rangle =\phi_{\alpha}(\vec{r}-\vec{R}_{j})$.}%
\footnote{We assume that all basis functions are real-valued and that $Y$ is
a local single-particle operator $\left\langle \vec{r}\right|Y\left|\vec{r}'\right\rangle =Y(\vec{r})\delta(\vec{r}-\vec{r}')$.
The overlap and Kohn-Sham fock operator of DFT are of this form.%
} The Gaussian basis functions used by TURBOMOLE are characterized
by the angular momentum $l$ and the multiplicity $\nu=1,\ldots,2l+1$,
and $\alpha$ is a collective index for both. The rotated basis functions
of angular momentum $l$ can be expressed as\cite{Hamermesh:Dover1989}
\[
\left[\psi'\right]_{\nu}^{l}(\vec{r})=\psi_{\nu}^{l}(\varrho^{-1}\vec{r})=\sum_{\mu=1}^{2l+1}\psi_{\mu}^{l}(\vec{r})D_{\mu\nu}^{l}(\varrho)\]
with the representation $D_{\mu\nu}^{l}(\varrho)$ of the rotation
$\varrho$. Using $Y'(\vec{r})=Y(\varrho^{-1}\vec{r})$, it can be
shown that the electrode parameters of the two coordinate systems
are related by\begin{equation}
Y_{\vec{R}_{j}\alpha,\vec{0}\beta}=\sum_{\mu,\nu}\left[D^{T}(\varrho)\right]_{\alpha\mu}Y'_{\varrho\vec{R}_{j}\mu,\vec{0}\nu}D(\varrho)_{\nu\beta},\label{eq:app_Ycluster_0ajb-rotation}\end{equation}
where $D(\varrho)$ is the representation of $\varrho$ in the employed
basis set. By knowledge of the $D_{\mu\nu}^{l}(\varrho)$, $D(\varrho)$
can be constructed by the process of the addition of representations.\cite{Hamermesh:Dover1989}
If there are $n_{l}$ basis functions of angular momentum $l$ in
the basis set describing $Y_{j\alpha,0\beta}$, we have\begin{equation}
D(\varrho)=\oplus_{l}n_{l}D^{l}(\varrho),\label{eq:electrodeDFT_addRep}\end{equation}
where $\oplus$ denotes a direct sum. 

Let us now give the explicit formulas for the $D^{l}(\varrho)$. In
this work only $s$, $p$, and $d$ basis functions are used, and
hence we restrict ourselves to $l=0$, $1$, and $2$. Since $s$
functions just depend on the radius, $\psi^{0}(\vec{r})=\psi^{0}(r)$,
we have \begin{equation}
D^{0}(\varrho)=1.\label{eq:electrodeDFT_D0_representation}\end{equation}
For $l=1$ there are three $p$ functions, $p_{1}=p_{x}=f_{1}(r)x$,
$p_{2}=p_{y}=f_{1}(r)y$, and $p_{3}=p_{z}=f_{1}(r)z$, with a certain
radial dependence $f_{1}(r)$.\cite{Atkins:Oxford1999} Exploiting
$\varrho^{-1}=\varrho^{T}$, we obtain $p'_{i}=\sum_{j=1}^{3}p_{j}\varrho_{ji}$.
Thus the $3\times3$ representation of the rotation $\varrho$ for
the $p$ functions is \begin{equation}
D^{1}(\varrho)=\varrho=\left(\begin{array}{ccc}
\varrho_{xx} & \varrho_{xy} & \varrho_{xz}\\
\varrho_{yx} & \varrho_{yy} & \varrho_{yz}\\
\varrho_{zx} & \varrho_{zy} & \varrho_{zz}\end{array}\right).\label{eq:electrodeDFT_D1_representation}\end{equation}
For $l=2$ there are five $d$ functions, where $d_{1}=d_{3z^{2}-r^{2}}=f_{2}(r)\left(3z^{2}-r^{2}\right)/\left(2\sqrt{3}\right)$,
$d_{2}=d_{xz}=f_{2}(r)xz$, $d_{3}=d_{yz}=f_{2}(r)yz$, $d_{4}=d_{xy}=f_{2}(r)xy$,
and $d_{5}=d_{x^{2}-y^{2}}=f_{2}(r)\left(x^{2}-y^{2}\right)/2$ with
some radial dependence $f_{2}(r)$. The transformed $d$ functions
are given as $d_{i}'=\sum_{j=1}^{3}d_{j}\left[D^{2}\left(\varrho\right)\right]_{ji}$
with the $5\times5$ representation {\small \begin{equation}
D^{2}(\varrho)=\left(\begin{array}{ccccc}
\left(3\varrho_{zz}^{2}-1\right)/2 & \sqrt{3}\varrho_{zx}\varrho_{zz} & \sqrt{3}\varrho_{zy}\varrho_{zz} & \sqrt{3}\varrho_{zx}\varrho_{zy} & \sqrt{3}\left(\varrho_{zx}^{2}-\varrho_{zy}^{2}\right)/2\\
\sqrt{3}\varrho_{xz}\varrho_{zz} & \varrho_{xx}\varrho_{zz}+\varrho_{zx}\varrho_{xz} & \varrho_{xy}\varrho_{zz}+\varrho_{xz}\varrho_{zy} & \varrho_{xx}\varrho_{zy}+\varrho_{xy}\varrho_{zx} & \varrho_{xx}\varrho_{zx}-\varrho_{xy}\varrho_{zy}\\
\sqrt{3}\varrho_{yz}\varrho_{zz} & \varrho_{yx}\varrho_{zz}+\varrho_{zx}\varrho_{yz} & \varrho_{yy}\varrho_{zz}+\varrho_{yz}\varrho_{zy} & \varrho_{yx}\varrho_{zy}+\varrho_{yy}\varrho_{zx} & \varrho_{yx}\varrho_{zx}-\varrho_{yy}\varrho_{zy}\\
\sqrt{3}\varrho_{xz}\varrho_{yz} & \varrho_{xx}\varrho_{yz}+\varrho_{yx}\varrho_{xz} & \varrho_{xy}\varrho_{yz}+\varrho_{yy}\varrho_{xz} & \varrho_{xx}\varrho_{yy}+\varrho_{xy}\varrho_{yx} & \varrho_{xx}\varrho_{yx}-\varrho_{xy}\varrho_{yy}\\
\sqrt{3}\left(\varrho_{xz}^{2}-\varrho_{yz}^{2}\right)/2 & \varrho_{xx}\varrho_{xz}-\varrho_{yx}\varrho_{yz} & \varrho_{xy}\varrho_{xz}-\varrho_{yy}\varrho_{yz} & \varrho_{xx}\varrho_{xy}-\varrho_{yx}\varrho_{yy} & \left(\varrho_{xx}^{2}+\varrho_{yy}^{2}-\varrho_{xy}^{2}-\varrho_{yx}^{2}\right)/2\end{array}\right).\label{eq:electrodeDFT_D2_representation}\end{equation}
}{\small \par}

\subsection{Imposing the fcc space-group\label{sec:electrodeDFT-impose-fcc}}

\begin{figure}
\begin{centering}\includegraphics[width=0.4\columnwidth,clip=]{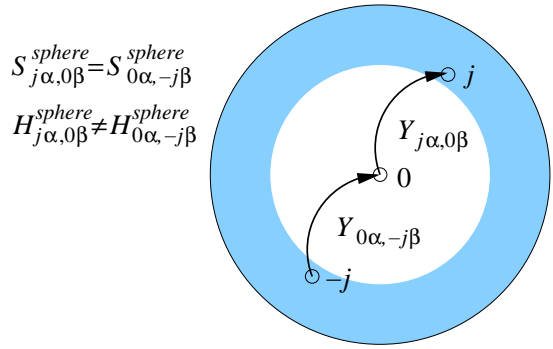}\par\end{centering}

\caption{\label{cap:electrodeDFT-HStrans}Necessity to impose the translational
symmetry on the hopping elements $H_{j\alpha,0\beta}^{sphere}$. A
finite spherical cluster is displayed. In blue-shaded areas surface
effects are important. While the overlap elements $S_{j\alpha,0\beta}^{sphere}$
are translationally symmetric, this is not the case for $H_{j\alpha,0\beta}^{sphere}$.
In particular $S_{j\alpha,0\beta}^{sphere}=S_{0\alpha,-j\beta}^{sphere}$,
while $H_{j\alpha,0\beta}^{sphere}\neq H_{0\alpha,-j\beta}^{sphere}$
as illustrated in the plot, where $Y=S,H$.}
\end{figure}

In this section we consider how to impose the fcc space-group on the
parameters $H_{j\alpha,0\beta}^{sphere}$, extracted from the finite
spherical fcc clusters {[}Fig.~\ref{cap:DFT-electrode-construction}].
Assuming basis functions to be real-valued, the matrix elements of
a translationally invariant Hamiltonian $H^{trans}$ are symmetric
and obey the relations \begin{equation}
H_{i\alpha,j\beta}^{trans}=H_{\left(i-j\right)\alpha,0\beta}^{trans}=H_{0\alpha,-\left(i-j\right)\beta}^{trans}=H_{-\left(i-j\right)\beta,0\alpha}^{trans}.\label{eq:electrodeDFT-Htrans_ij}\end{equation}
Owing to surface effects, the translational symmetry is not fulfilled
by the parameters $H_{j\alpha,0\beta}^{sphere}$, as illustrated in
Fig.~\ref{cap:electrodeDFT-HStrans}.%
 Hence, although the deviations decrease with growing radius of the
spheres, the translational symmetry needs to be enforced in order
to describe a crystal. To avoid numerical errors, we impose at the
same time the point-group symmetry $O_{h}$ although that symmetry
is already present due to the shape of our clusters. Concerning the
notation, we will call the parameters conforming to the $O_{h}$ point-group,
the translational symmetry, and the fcc space-group $H_{j\alpha,0\beta}^{O_{h}}$,
$H_{j\alpha,0\beta}^{trans}$, and $H_{j\alpha,0\beta}=H_{j\alpha,0\beta}^{fcc}$,
respectively. We do not need to consider the overlap, since it depends
only on the relative position of two atoms.

\paragraph*{$O_{h}$ point-group symmetry}

With Eq.~(\ref{eq:app_Ycluster_0ajb-rotation}) a Hamiltonian $H_{j\alpha,0\beta}^{O_{h}}$
conforming to the point-group symmetry can be constructed by averaging,
for a given element of $H_{j\alpha,0\beta}^{O_{h}}$, over all $H_{j\alpha,0\beta}^{sphere}$
related to it by symmetry\begin{equation}
H_{\vec{R}_{j}\alpha,\vec{0}\beta}^{O_{h}}=\frac{1}{N_{O_{h}}}\sum_{\varrho\in O_{h}}\sum_{\mu,\nu}\left[D^{T}(\varrho)\right]_{\alpha,\mu}H_{\varrho\vec{R_{j}}\mu,\vec{0}\nu}^{sphere}D(\varrho)_{\nu\beta}.\label{eq:electrodeDFT_impose_point_symm}\end{equation}
Here, $\varrho$ runs over all $N_{O_{h}}=48$ symmetry elements of
the point group $O_{h}$.\cite{Hamermesh:Dover1989,DalCorso:Berlin1996}

\paragraph*{Translational symmetry}

Using Eq.~(\ref{eq:electrodeDFT-Htrans_ij}), the translational symmetry
can be imposed by setting\begin{equation}
H_{j\alpha,0\beta}^{trans}=\frac{1}{2}\left(H_{j\alpha,0\beta}^{sphere}+H_{-j\beta,0\alpha}^{sphere}\right).\label{eq:electrodeDFT-Htrans}\end{equation}

\paragraph*{Fcc space-group}

The combined action of the $O_{h}$ point-group and the translational
symmetry leads to the fcc space-group symmetry. With Eqs.~(\ref{eq:electrodeDFT_impose_point_symm})
and (\ref{eq:electrodeDFT-Htrans}) we obtain the fcc space-group
symmetric parameters $H_{j\alpha,0\beta}=H_{j\alpha,0\beta}^{fcc}$
according to the prescription\begin{equation}
H_{\vec{R_{j}}\alpha,\vec{0}\beta}^{fcc}=\frac{1}{2N_{O_{h}}}\sum_{\varrho\in O_{h}}\sum_{\mu,\nu}\left\{ \left[D^{T}\left(\varrho\right)\right]_{\alpha\mu}H_{\varrho\vec{R_{j}}\mu,\vec{0}\nu}^{sphere}D\left(\varrho\right)_{\nu\beta}+\left[D^{T}\left(\varrho\right)\right]_{\beta\mu}H_{-\varrho\vec{R_{j}}\mu,\vec{0}\nu}^{sphere}D\left(\varrho\right)_{\nu\alpha}\right\} .\label{eq:electrodeDFT-Hfcc}\end{equation}

\end{document}